# Precise Photon Correlation Measurement of a Chaotic Laser

**Xiaomin Guo[1,2], Chen Cheng[1,2], Tong Liu[1,2], Xin Fang[1,2] and Yanqiang Guo[1,2],***

[1] Key Laboratory of Advanced Transducers and Intelligent Control System, Ministry of Education, Taiyuan University of Technology, Taiyuan 030024, China; guoxiaomin@tyut.edu.cn (X.G.); chengchen248@163.com (C.C.); liutong0912@163.com (T.L.); fangxin0837@126.com (X.F.)

[2] College of Physics and Optoelectronics, Taiyuan University of Technology, Taiyuan 030024, China;

\* Correspondence: guoyanqiang@tyut.edu.cn (Y.G.)

**Featured Application: This technique of improving the accuracy of $g^{(2)}(\tau)$ measurement is useful to extract higher order coherence and achieve desired laser source for quantum imaging and secure communication.**

**Abstract:** The second order photon correlation $g^{(2)}(\tau)$ of a chaotic optical-feedback semiconductor laser is precisely measured using Hanbury Brown-Twiss interferometer. The accurate $g^{(2)}(\tau)$ with non-zero delay time is obtained experimentally from the photon pair time interval distribution through a ninth-order self-convolution correction. The experimental results agrees well with the theoretical analysis. The relative error of $g^{(2)}(\tau)$ is no more than 5‰ within 50 ns delay time. The bunching effect and coherence time of chaotic laser are measured via the precise photon correlation technique. This technique provides a new tool to improve the accuracy of $g^{(2)}(\tau)$ measurement and boost the applications of quantum statistics and correlation.

## 1. Introduction

Semiconductor lasers subject to external optical feedback exhibit a rich variety of nonlinear dynamical behaviors and are used to generate high-dimensional chaotic laser [1,2]. This configuration has attracted great interest for a wide range of applications, like optical chaos communication [3-7], secure key distribution [8], high-speed physical random number generation [9-12], chaos-based optical computing [13] and sensing [14-16]. It is fundamentally important to understand the underlying physical mechanisms of chaotic laser, and practically useful to improve the laser performance and motivate its applications. Previous researches mainly focused on clarifying intensity statistics and autocorrelation (AC) of chaotic lasers to characterize the chaotic processes [17-21].Intensity statistics are closely relevant to the extractible rate of random number [9,10,22] and the AC is a good indicator of chaotic modulating bandwidth in optical chaos communications [4,19]. However, the macro-scale intensity statistics and AC are not sufficient to reveal all properties of a given chaotic laser, and there is also a significant discrepancy between experimental and theoretical probability density distributions of the laser intensity [19]. Recent researches reveal that quantum correlation is more accurate in assessing statistical properties and more sensitive to control parameters compared to AC function [23,24]. However, the previous research is concentrated on the properties of the quantum dot laser in the low-intensity (low-gain) situation, and the bunching effect of the chaotic laser, i.e., $g^{(2)}(0) > 1$ at $\tau = 0$, is only revealed in the fully developed chaotic (high-gain) regime. Studies on high order photon correlation of high-dimensional chaotic lasers are sparse, especially second order photon correlation $g^{(2)}(\tau)$ at non-zero delay time.

The landmark experiment on photon correlation was first conducted by Hanbury Brown and Twiss (HBT), demonstrating spatial second order photon correlation $g^{(2)}$ of a thermal light [25]. Soon afterwards, this experiment inspired Glauber's seminal work on quantum optics theory, which described the photon correlation of different light fields by correlation functions within quantum statistics [26-28]. The photon correlation $g^{(2)}$ is fundamentally different from the first order correlation and is harnessed in many applications, such as photon bunching and anti-bunching measurement [29-32], spatial interference [33,34], ghost imaging [35-37], the azimuthal HBT effect [38], single photon detection [39], etc. The $g^{(2)}(\tau)$ also carries a wealth of information on the statistical probability of different photons arriving at time delay $\tau$ [40]. Up until now, there are many approaches to obtain the photon correlation $g^{(2)}(\tau)$, typically including two-photon absorption (TPA) measurement [41,42], photon coincidence counting [43], time interval measurement of photon pairs [44]. Recently, the HBT experiment was explored to observe chaos from quantum dot lasers with external feedback [23]. However, research on photon correlation $g^{(2)}(\tau)$ of high-dimensional chaotic waveforms is rare and there is still an obvious disagreement between experimental and theoretical $g^{(2)}(\tau)$. The calculation of $g^{(2)}(\tau)$ from the photon pair time interval distribution provides a good way to measure the photon correlation of pseudo-thermal light with microsecond coherence time [44]. But For chaotic laser the coherence time is much shorter than that of pseudo-thermal light, and the resolution time must be shorter than the coherence time of the laser in the measurement. Although the shorter coherence time does not affect the bunching effect or $g^{(2)}(0)$ of chaotic laser [24], that makes the measurement of $g^{(2)}(\tau)(\tau\neq0)$ at very short timescales using HBT technique more difficult, owing to the limited response time of single-photon detectors [45]. It remains an important challenge to ravel the $g^{(2)}(\tau)$ ($\tau\neq0$) of the chaotic laser at a high precision, whose coherence time is below 1 ns. Accordingly, high precision and ultrashort resolution time are required to acquire an accurate $g^{(2)}(\tau)$ of the chaotic laser. That is, it is potentially useful to extract higher order coherence and achieve desired laser source for quantum imaging and secure communication.

In this paper, we theoretically and experimentally investigate the second order photon correlation $g^{(2)}(\tau)$ of a chaotic optical-feedback semiconductor laser. The $g^{(2)}(\tau)$ is precisely measured using self-convolution HBT detection at tens of picoseconds resolution time. A different high order correction of $g^{(2)}(\tau)$ is analyzed and confirmed experimentally, which has a low relative error in wide range of delay time. It shows a good agreement between experimental results and theoretical analysis. We also measure the bunching effect and coherence time of chaotic laser using the precise photon correlation technique. This technique, avoiding the photon overlapping, can give a $g^{(2)}(\tau)$ with a high accuracy. To the best of our knowledge, the accurate measurement of $g^{(2)}(\tau)$ for the chaotic laser has not been investigated and reported. In view of this demonstration, we present first some highlights of precise photon correlation measurement that are necessary for a better understanding of quantum statistics of the chaotic laser. The demonstration well reveals photon correlation $g^{(2)}(\tau)$ of chaotic laser and provide a way of studying chaos with quantum optics technique.

**2. High Order Correction of $g^{(2)}(\tau)$**

Theoretically, second order photon correlation of $g^{(2)}(\tau)$ can be obtained from an ideal photon pair time interval distribution $P_1(\tau)$. Using the self-convolution method, one can obtain any desired high order $n$, and the higher $n$ of $g_n^{(2)}(\tau)$ is, the more accurate $g_n^{(2)}(\tau)$ tends to the ideal $g^{(2)}(\tau)$. But due to the actual operation capacity of data processing and the difficulty of convolving complex form to very high order, we reasonably convolve $g_n^{(2)}(\tau)$ to the ninth order but the relative error is small enough to obtain high accuracy.

In our experiment, photon pair time interval distribution is collected by single photon counters and the time distribution is $D_1(\tau)$. Furthermore, $g^{(2)}(\tau)$ can be calculated from the self-convolution of $D_1(\tau)$.

The second order photon correlation $g^{(2)}(\tau)$ have a proportional relation to $G(\tau)$, as follows:

$$G(\tau) = \bar{I} g^{(2)}(\tau), \tag{1}$$

where $\bar{I}$ is the average photon counting rate per time bin of the light field. $G(\tau)$ is the histogram of photons at delay time $\tau$ between two photon detection events. The relation between $G(\tau)$ and $P_1(\tau)$ is given by,

$$G(\tau) = P_1(\tau) + P_1(\tau) * P_1(\tau) + \ldots\ldots = \sum_{n=1}^{\infty} P_n(\tau), \tag{2}$$

where $P_1(\tau)$ is an ideal photon pair time interval distribution of light field which can be obtain based on HBT experiment, and $P_n(\tau)$ is $n$th order self-convolution of $P_1(\tau)$. When $P_1(\tau)$ is less than one, the sum of $P_n(\tau)$ is convergence [40], then we can obtain

$$P_1(\tau) = L^{-1}\left(\frac{L(G(\tau))}{1+L(G(\tau))}\right), \tag{3}$$

where $L$ denotes the Laplace transformation, and $L^{-1}$ denotes the Inverse Laplace transformation.

When the above theory is applied to Lorentzian chaotic laser field, we can get the relation between $G(\tau)$ and $P_n(\tau)$ of a chaotic laser. The first order correlation of Lorentzian chaotic laser is as follows:

$$g^{(1)}(\tau) = e^{-\frac{|\tau|}{\tau_c}}. \tag{4}$$

The relation between $g^{(2)}(\tau)$ and $g^{(1)}(\tau)$ of Lorentzian chaotic laser is:

$$g^{(2)}(\tau) = 1 + \left|g^{(1)}(\tau)\right|^2. \tag{5}$$

Using Equations (4) and (5), we obtain:

$$g^{(2)}(\tau) = 1 + \left|e^{-|\tau|/\tau_c}\right|^2. \tag{6}$$

According to Equations (1) and (6), we obtain:

$$G(\tau) = \bar{I}(1 + \left|e^{-|\tau|/\tau_c}\right|^2). \tag{7}$$

The relation between $g^{(2)}(\tau)$ and $P_n(\tau)$ is shown as follows.

$P_2(\tau)$ is the self-convolution of $P_1(\tau)$

$$P_2(\tau) = \int_0^{+\infty} P_1(\tau) P_1(t-\tau) dt = P_1(\tau) * P_1(\tau). \tag{8}$$

$P_3(\tau)$ is the convolution of $P_1(\tau)$ and $P_2(\tau)$

$$P_3(\tau) = \int_0^{+\infty} P_2(\tau) P_1(t-\tau) dt = P_2(\tau) * P_1(\tau). \tag{9}$$

$P_n(\tau)$ is the convolution of $P_1(\tau)$ and $P_{n-1}(\tau)$

$$P_n(\tau) = \int_0^{+\infty} P_{n-1}(\tau) P_1(t-\tau) dt = P_{n-1}(\tau) * P_1(\tau). \tag{10}$$

The integration upper bound of Equation (10) should be replaced by the maximum time interval $\tau$ practically. Now the new equation is

$$P_n(\tau) = \int_0^{\tau} P_{n-1}(\tau - t) P_1(\tau) dt = P_{n-1}(\tau) * P_1(\tau). \tag{11}$$

Using Equation (3), (8), (9), (10), and (11), we obtain all of the self-convolution of $P_1(\tau)$

Then we obtain the following equations

$$P_2(\tau) = L^{-1}(\frac{L(G(\tau))}{1+L(G(\tau))})*L^{-1}(\frac{L(G(\tau))}{1+L(G(\tau))}) \tag{12}$$

$$P_n(\tau) = \overbrace{L^{-1}(\frac{L(G(\tau))}{1+L(G(\tau))})*\cdots*L^{-1}(\frac{L(G(\tau))}{1+L(G(\tau))})}^{N} \tag{13}$$

$$\begin{aligned}g_n^{(2)}(\tau) = & (L^{-1}(\frac{L(\bar{I}g^{(2)}(\tau))}{1+L(\bar{I}g^{(2)}(\tau))}) \\ & + L^{-1}(\frac{L(\bar{I}g^{(2)}(\tau))}{1+L(\bar{I}g^{(2)}(\tau))})*L^{-1}(\frac{L(\bar{I}g^{(2)}(\tau))}{1+L(\bar{I}g^{(2)}(\tau))})+\cdots \\ & + \overbrace{L^{-1}(\frac{L(\bar{I}g^{(2)}(\tau))}{1+L(\bar{I}g^{(2)}(\tau))})*\cdots*L^{-1}(\frac{L(\bar{I}g^{(2)}(\tau))}{1+L(\bar{I}g^{(2)}(\tau))})}^{N})/\bar{I}\end{aligned} \tag{14}$$

Inserting Equation (7) to Equation (13) we can get different $P_n(\tau)$. The form of $P_n(\tau)$ can be obtained by numerical self-convolution. The sum of $P_n(\tau)$ is $G(\tau)$, and in theory $g_n^{(2)}(\tau)$ is comparable to $g^{(2)}(\tau)$ for sufficiently high n. In fact, with the increase of $n$, $g_n^{(2)}(\tau)$ is closer to ideal $g^{(2)}(\tau)$. Using Equation (14) and increasing the order of $n$, we can obtain high order $g_n^{(2)}(\tau)$. Considering the realistic experiment condition and the data-processing ability, the maximum order of $n$ we take is nine. The theoretical high order correction of $g^{(2)}(\tau)$ is given above, which can help us to know the influences of the experimental parameters. Here, the direct self-convolution method is used to get $g_n^{(2)}(\tau)$ from experimental data. In that case, $P_1(\tau)$ is related to the experimentally measured photon pair time interval distribution $D_1(\tau)$. $D_n(\tau)$ is $n$th order self-convolution of $D_1(\tau)$. Experimental results of $g_n^{(2)}(\tau)$ can be obtained from $D_1(\tau)$ [44],

$$D_1(\tau) = \sum_{n=1}^{\infty}\frac{1}{2^n}P_n(\tau), \tag{15}$$

and the relation between $G(\tau)$ and $D_n(\tau)$ is

$$G(\tau) = 2\sum_{n=1}^{\infty}D_n(\tau). \tag{16}$$

Thus, when we obtain the $D_1(\tau)$, the high order correction $g_n^{(2)}(\tau)$ can be deduced from the experimental photon pair time interval distribution as follows

$$g_n^{(2)}(\tau) = \frac{1}{\bar{I}}\cdot 2\sum_{n=1}^{\infty}D_n(\tau). \tag{17}$$

The above analysis basically solves the high order correction $g_n^{(2)}(\tau)$ of the chaotic laser in theory and experiment. One can also use this method to analyze the error caused by the variations of the mean photon intensity and coherence time of chaotic laser. In addition, high order correction of $g^{(2)}(\tau)$ for coherent light can be achieved and the $g^{(2)}(\tau)$ is perfectly equal to one.

Using the Equation (14), $g_n^{(2)}(\tau)$ is calculated to ninth order, and higher order terms than ninth can be omitted. Relative error δ varying with the delay time τ at the correction order of nine is calculated as:

$$\delta = \frac{\left|g_n^{(2)}(\tau) - g^{(2)}(\tau)\right|}{g^{(2)}(\tau)} \times 100\% .\qquad(18)$$

**3. Experiment Setup**

The experimental setup is shown in Figure 1, which can be used to determine time and frequency domain of the laser characteristics and measure photon pair time interval distribution. A 1550 nm laser is generated by a distributed feedback laser diode (DFB-LD), and a thermoelectric temperature controller (TTC, ILX-Light wave LDT-5412) was used to stabilize the temperature with an accuracy of 0.01 K. A precision current source controller (CSC, ILX-Light wave LDC-3412) controlled the output intensity of the DFB-LD laser nearby 1.5 times threshold current with a value of 15.9 mA. The output laser passes through a polarization controller (PC) which maintains the polarization of the feedback beam paralleling to that of the output laser. With the help of an optical circulator (OC), the optical feedback loop can be realized. The output of the OC is connected to a 20:80 fiber coupler (FC). A total of 80% of the output light passed through the variable optical fiber attenuator (VA1) and went back to the OC. Another output is connected with a 50:50 FC, and one port of output was detected by a high-speed photo detector (PD, FINISAR XPDV2120RA). Signal time series were recorded by an oscilloscope (OSC, Lecroy LabMaster10-36Zi) and the frequency spectrum was obtained by a frequency spectrum analyzer (Agilent N9020A). On the same output port of this 50:50 FC, the optical spectrum was also measured by an optical spectrum analyzer (Yokogawa AQ6370C). The other output port of this 50:50 FC was connected with another attenuator (VA2) followed by an HBT system, which was based on a 50:50 beam splitter (BS) with a dual channel single photon detector (SPD, Aurea Technology LYNXEA. NTR. M2). When the photons impinge on the SPD, the SPD delivers pulses to a time to digital converter (TDC). An internal clock triggered two channel gates simultaneously. Then precise time information (i.e., the time between photons arrival at different channels) was extracted via a subtractor and an integrator. Each photon pair time interval was placed in the one-time bin. The histogram of the photon pair time interval distribution is obtained through cumulative measurement. The data ware read out to a laptop computer (LC) via universal serial bus (USB) connection. When the laser beam passes from the fiber to space or space to fiber, the fiber lens collimators are required. In Figure 1, L1, L2, L3 represents the aspheric lens collimators, and F is an optical filter used to filter out the background noise. The chaotic laser is divided into two equal intensity beams whose intensity are measured by the detectors SPD1 and SPD2. One can adjust the mean photon intensity of the light through the VA2. After the above steps, the photon pair time interval distribution can be attained.

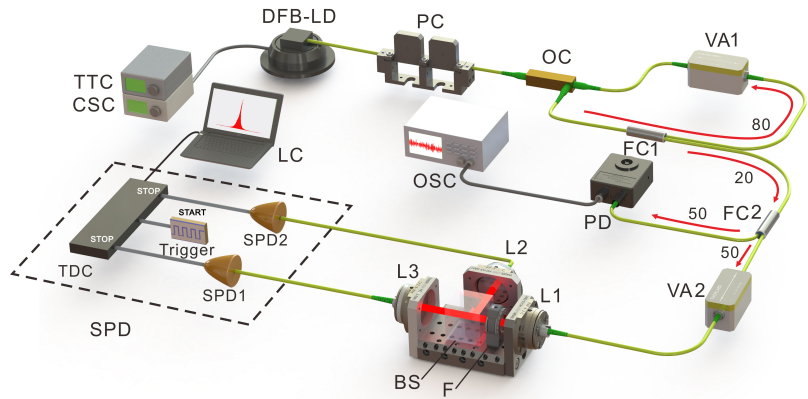

**Figure 1**. Schematic diagram of the experimental setup for measuring photon correlation $g^{(2)}(\tau)$ of a chaotic optical-feedback laser. DFB-LD: distributed feedback 1550 nm laser diode; PC: polarization controller; OC: optical coupler; VA1 and VA2: variable optical fiber attenuator; FC1 and FC2: fiber coupler; PD: high-speed photo detector; L1, L2 and L3: aspheric lens collimator; BS: beam splitter; F: optical filter; SPD: single photon detector; TDC: time to digital converter; LC: laptop computer; OSC: oscilloscope; TTC: thermoelectric temperature controller; CSC: current source controller.

## 4. Experimental Results

The chaotic laser is firstly attenuated by a variable attenuator and then passes through the HBT setup. In the photon detection system, an internal clock triggers two channel gates simultaneously. When a photon was detected on one channel, the arriving time is recorded. During the same clock period, a subsequent photon was received from another channel and then the time interval was measured. The desired distribution was obtained with many records, and if the detection quantum efficiency is higher, the better the photon pair distribution is close to the real light source distribution. Otherwise, the single photon detector would mistake dark noise for photon signals. Moreover, as the incident photon number increases, the noise level would be higher due to the after pulsing effect. In that case, the time interval distribution of photon pairs is also affected by noise. When the coherence time of light source is short, high resolution time is required in the detection. Besides, the unbalance of the two light intensity after the BS has an adverse effect on the acquired distribution. It was difficult to obtain an accurate time interval distribution of photon pairs with a very low quantum efficiency. In our experiment, the detection quantum efficiency is 25%. We investigate different average photon intensity and coherence time affect the accuracy of different order correction. We use the relative error to compare different high order correction with the ideal second order photon correlation. According to Equation (2), we calculate $P_9(\tau)$ with high order terms and omitted the terms higher than ninth order. Likewise, we take photon pair time interval distribution $D_1(\tau)$ and then convolve $D_1(\tau)$ to $D_9(\tau)$. The terms higher than ninth order is also omitted. Using Equation (17) we obtain different high order correction of $g^{(2)}(\tau)$ with experiment data. The influence of different average photon intensity and the coherence time are investigated theoretically.

At 1.5 times the threshold current ($J=1.5J_{th}$) and 25 °C temperature (T=25°C), central wavelength was stabilized near 1548 nm. We adjust the attenuator VA1 and polarization controller to accurately control the optical feedback strength. With the increase of the feedback strength, the laser experienced a transition from the period-1, period-2, to the steady chaos oscillation. Among them, we select three typical states, including period-2 (weak chaos) with the feedback strength $\eta$ of 2.66%, the intermediate chaotic state (chaos) with $\eta$ of 8.87%, and steady chaotic oscillation state (Strong chaos) with $\eta$ of 30.31%. Figure 2a shows the three typical frequency spectrums of the chaotic laser. To quantify the bandwidth of the chaotic laser, we used the definition that is expounded as the frequency spectrum region the DC and the frequency where 80% of the energy is contained within [46]. According to the 80% bandwidth definition, the bandwidth of chaotic laser was 4.98 GHz, 9.84 GHz, and 11.71 GHz, respectively. Figure 2b is the optical spectrum of the chaotic laser. Environmental changes slightly influence the optical feedback strength and the coherence length [47]. Based on the repeated measurements we obtained the range of coherent time variation. Figure 3 is the three corresponding time series of the chaotic laser.

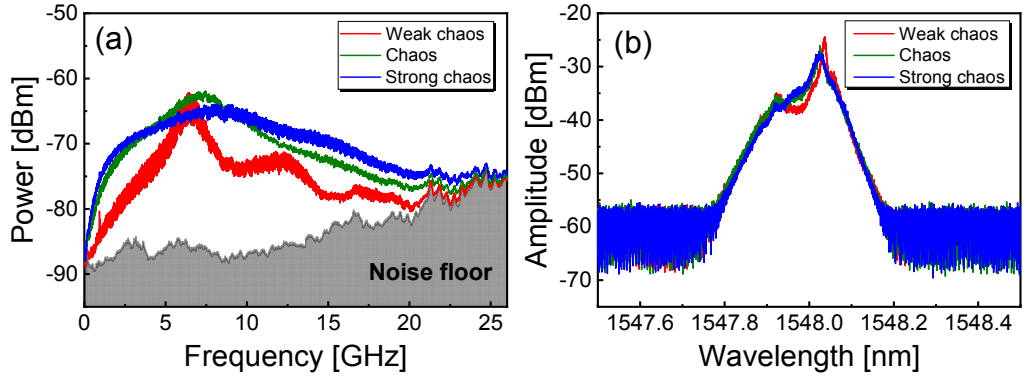

**Figure 2.** (a) Measured frequency spectrum and (b) optical spectrum of the chaotic laser, when J = 1.5Jth and η = 2.66% (weak chaos), 8.87% (chaos), 30.31% (strong chaos).

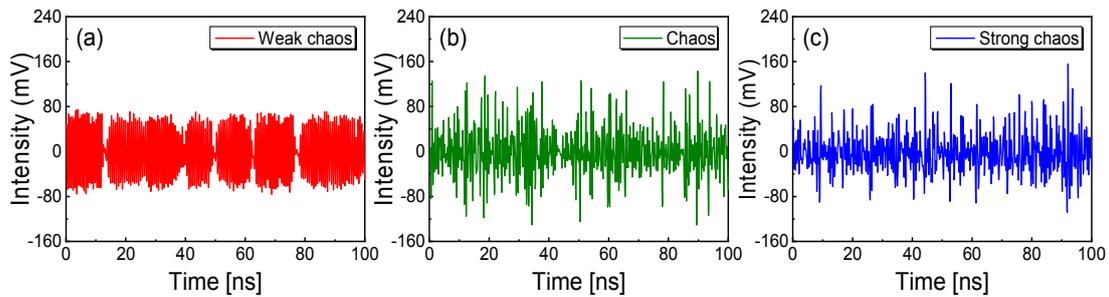

**Figure 3.** Measured three corresponding time series of the chaotic laser. The parameters of bias current and feedback strength are the same as those used in Figure 2.

The bandwidth of the chaotic laser is in the order of GHz and we obtained the coherence time of chaotic laser through 3dB linewidth spectrum. Considering that the ninth order correction of the second order photon correlation $g^{(2)}(\tau)$ was close enough to theoretical limit, we experimentally took the ninth order correction within 10 ns and theoretically employ the same order fitting. The experimental photon correlation $g^{(2)}(\tau)$ are fitted by ideal expressions, as shown in Figure 4. For photon-bunching chaotic light, the $g^{(2)}(\tau)$ can be written as $g^{(2)}(\tau)=1+b\exp(-2\tau/\tau_c)$ (*b*: bunching amplitude, $\tau_c$: coherence time) [23]. Figure 4 shows the experimental and theoretical fitting results for weak chaos (*b*=0.479, $\tau_c$=0.768 ns), chaos (*b*=0.524, $\tau_c$=0.651 ns), and strong chaos (*b*=0.626, $\tau_c$=0.535 ns).

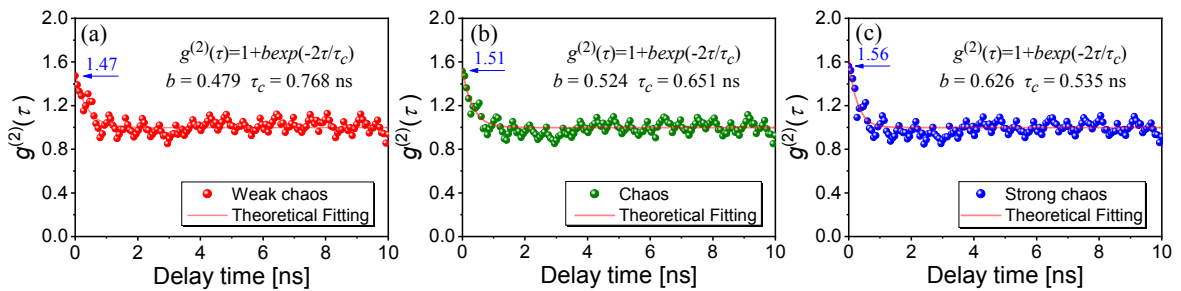

**Figure 4.** The 9th order correction of second order photon correlation and the theoretical fitting. The bias current and feedback strength are the same as those in Figure 2.

## 5. Influences of Detector Time Resolution and High Order Omitted Terms

In our experiment, the resolution time of the detection (65 ps) was not significantly small compared to the coherence time (~0.5 ns) of the chaotic laser, resulting in a little fluctuation of measured $g^{(2)}(\tau)$. In Figure 5(a), the experimental results of $g_n^{(2)}(\tau)$ within 100 ns delay time is shown and the magenta curve represent the original photon pair time interval distribution. The original experimental data is the same as those used in Fig 4b. The bottom-up colored curves indicate the increasing order corrections of second order photon correlation. The orange curve indicate is the third order correction of $g^{(2)}(\tau)$, and the others are fifth, seventh, ninth order correction of $g^{(2)}(\tau)$. For an accurate measurement of photon correlation, a very low photon flux rate I was required to ensure $I\tau_0 \ll 1$[39]. The courting rate of the SPD was controlled below 0.3 Mcounts/s by using the VA2 and the overall detection efficiency was 25%. In figure 5a, the counting rate was 270 kcounts/s and the dead time of the SPD was 4 us. Within 100 ns sampling time (I.e., τ=100ns), the incident light intensity was estimated to be about $4 \times 10^7$ photons/s. Figure 5b shows the theoretical results when $\tau_0$ was 0.5 ns and the incident light intensity was $4 \times 10^7$ photons/s. The experimental results are in good agreement with the theory.

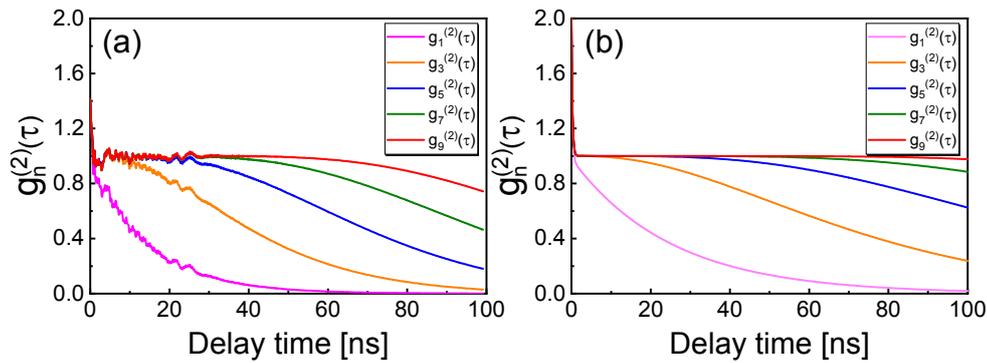

**Figure 5.** The **(a)** experimental and **(b)** theoretical results of $g_n^{(2)}(\tau)$ within the delay time 100ns.

## 6. Relative Error of $g_n^{(2)}(\tau)$ With Mean Photon Intensity and Coherence Time of Chaotic Laser

The coherence time of our experiment is below 1 ns and we can set the maximum coherence time in the theoretical analysis. Following this $g^{(2)}(\tau)$ is obtained by using Equation (6), which is independent on the mean photon intensity. Furthermore, according to Equation (18), it is found that the mean photon intensity and coherence time have effects on the relative error. The maximum photon intensity in our experiment is not exceed 0.05 photons/ns. Given this finding, we changed the mean photon intensity from 0.03 photons/ns to 0.05 photons/ns. For the low order correction of $g^{(2)}(\tau)$, it cannot provide sufficient information and accuracy according to Equation (2). For the ninth order correction, there was almost no difference between $g_9^{(2)}(\tau)$ and $g^{(2)}(\tau)$ and the loss information can be ignored. Figure 6 shows relative error of $g_9^{(2)}(\tau)$ for photon intensity changes form 0.03 photons/ns to 0.05 photons/ns and different delay times with the ninth order correction. The relative error varied with the photon intensity and delay time. When the delay time is shorter than 40ns the relative error can be ignored, while the relative error is increased when the delay time is close to 100 ns. It should be noted that higher order correction can reduce the relative error for longer delay time. In Figure6, it is also indicated that larger photon intensity brings bigger error. But when the photon intensity is too low, the photon pair time interval distribution contains a lot of dark noise that deteriorates the detection performance.

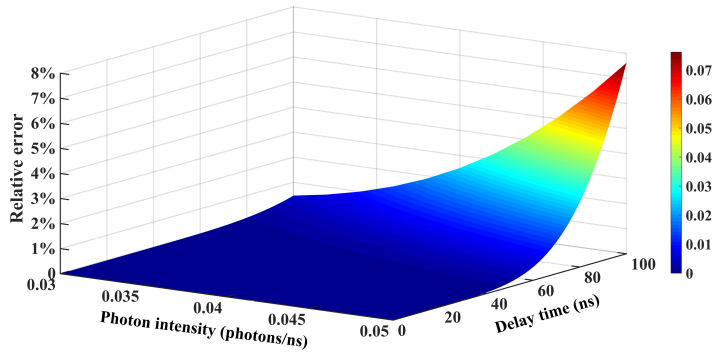

**Figure 6.** The relative error varying with the photon intensity and delay time. The coherence time is set as 1ns.

Following this, we theoretically analyzed the coherence time from 0.3 ns to 0.7 ns under the condition that the photon intensity is near 4×10⁷ photon/sec. Figure 7 shows the relative error as functions of the coherence time $τ_c$ and the delay time $τ$. The coherence time $τ_c$ varies from 0.3 ns to 0.7 ns and the delay time $τ$ is within 100 ns. In this case, corresponding to our experimental condition, the relative error is not exceed 5‰ within 50 ns delay time. It is worth noting that long $τ_c$ leads to big relative error, but the change of relative error is subtle. The relative error caused by the coherence time is smaller than that of the photon intensity.

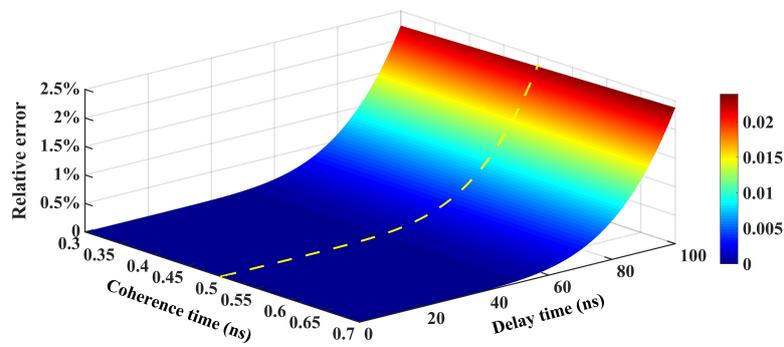

**Figure 7.** Relative error as functions of coherence time and delay time. The photon intensity is 0.04 photons/sec and the yellow dashed line corresponds to the experimental condition.

We compare the relative error caused by the above two factors (photon intensity and coherence time). The yellow dashed line in Figure 7 indicates the case that the coherence time is 0.5 ns, which corresponds to the experiment condition. For the same delay time, the relative error caused by coherence time was lower than that caused by photon intensity. Thus, high accuracy $g^{(2)}(τ)$ requires well controlling the photon intensity [24].

From the above discussion, the high order correction of second order photon correlation was affected by the variations of the mean photon intensity and coherence time of the laser, and we analyzed the relative error caused by the two factors respectively. The relative error from incident photon intensity was larger than that from coherence time. In Figure 7, the dashed line on the error surface was under the condition that the intensity was 0.04 photons/ns and $τ_c$ was 0.5 ns, which corresponds to the experimental condition. In our experiment, the maximum relative error in ninth order correction of $g^{(2)}(\tau)$ does not exceed 5‰ within 50 ns delay time. The relative errors caused by the photon intensity and coherence time retained the uncertainty ±0.01 photon/ns and ±0.2 ns respectively, and the overall error within 50 ns delay time did not exceed 1% in our condition.

## 8. Conclusions

In conclusion, we precisely measured the second order photon correlation $g^{(2)}(\tau)$ of a chaotic semiconductor laser using self-convolution HBT interferometer. Based on the theoretical analysis, the ninth order self-convolution correction was sufficient to obtain experimentally the accurate $g^{(2)}(\tau)$ from the photon pair time interval distribution. The experimental results were in good agreement with the theory. The relative error caused by coherence time and mean photon intensity was analyzed, which was no more than 5‰ within 50 ns delay time. In comparison with the traditional HBT measurement, this technique, which does not require high intensity and long optical or electric delay, is more useful for a weak light source, such as atomic fluorescence and single photon emission, whose quantum correlation is difficult to be detected. It is demonstrated that this technique provides a new way to measure high order quantum coherence precisely and will bridge the gap between nonlinear optics of chaotic lasers and quantum physics.

**Author Contributions:** X.G. and Y.G. designed the whole work and wrote the manuscript; X.G. carried out the theoretical calculations and analyzed the data; Y.G. supervised the experiments; C.C. and T.L. contributed to the experiment and data processing; X.F. analyzed the data and edited the manuscript. All authors discussed the results at all stages. All authors have read and approved the final manuscript.

**Funding:** This research was funded by the National Natural Science Foundation of China (NSFC) (Grants Nos. 61875147, 61671316, 61705160, 61731014), the Shanxi Scholarship Council of China (SXSCC) (Grant No. 2017-040), the Natural Science Foundation of Shanxi Province (Grants Nos. 201701D221116, 201801D221182), the Scientific and Technological Innovation Programs of Higher Education Institutions in Shanxi (STIP) (Grant No. 201802053), and the Program of State Key Laboratory of Quantum Optics and Quantum Optics Devices (Grant No. KF201905).

**Acknowledgments:** The authors thank Yi-Wei Liu and Chen-How Huang for helpful discussions.

**Conflicts of Interest:** The authors declare no conflict of interest.